\documentclass[preprint,3p,times,twocolumn]{elsarticle}
\usepackage{color}% Include figure files
\usepackage{epsfig}
\usepackage{amssymb}
\usepackage{amsmath}
\usepackage{dsfont}
\biboptions{sort&compress}

\usepackage{graphicx,hyperref}% Include figure files
\usepackage{bm}% bold math
\usepackage{subfigure}% bold math

\newcommand{\beq}{\begin{equation}}
\newcommand{\eeq}{\end{equation}}
\newcommand{\ba}{\begin{array}}
\newcommand{\bea}{\begin{eqnarray}}
\newcommand{\ea}{\end{array}}
\newcommand{\eea}{\end{eqnarray}}

\newcommand\eqn[1]{(\ref{#1})}      % parentheses around the LaTex "ref" macro
\newcommand\Eqn[1]{Eq.~(\ref{#1})}  % includes ``Eq.'' in front
\newcommand{\bX}{{\bf X}}

\newcommand{\C}{\mathcal{C}}

\newcommand{\bce}{\begin{center}}
\newcommand{\ece}{\end{center}}

\journal{Physics Letters B}

\begin{document}

\begin{frontmatter}

%% Title, authors and addresses

%% use the tnoteref command within \title for footnotes;
%% use the tnotetext command for the associated footnote;
%% use the fnref command within \author or \address for footnotes;
%% use the fntext command for the associated footnote;
%% use the corref command within \author for corresponding author footnotes;
%% use the cortext command for the associated footnote;
%% use the ead command for the email address,
%% and the form \ead[url] for the home page:
%%
%% \title{Title\tnoteref{label1}}
%% \tnotetext[label1]{}
%% \author{Name\corref{cor1}\fnref{label2}}
%% \ead{email address}
%% \ead[url]{home page}
%% \fntext[label2]{}
%% \cortext[cor1]{}
%% \address{Address\fnref{label3}}
%% \fntext[label3]{}

\title{Renormalization group flow and symmetry restoration in de Sitter space}

%% use optional labels to link authors explicitly to addresses:
%% \author[label1,label2]{<author name>}
%% \address[label1]{<address>}
%% \address[label2]{<address>}

\author{J. Serreau \corref{cor1}} \ead{serreau@apc.univ-paris7.fr}%
\address{APC, AstroParticule et Cosmologie, Universit\'e Paris Diderot, CNRS/IN2P3, CEA/Irfu, Observatoire de Paris, Sorbonne Paris Cit\'e,\\ 10, rue Alice Domon et L\'eonie Duquet, 75205 Paris Cedex 13, France}

\begin{abstract}
We compute the renormalization group flow of $O(N)$ scalar field theories in de Sitter space using nonperturbative renormalization group techniques in the local potential approximation. We obtain the flow of the effective potential on superhorizon scales for arbitrary space-time dimension $D=d+1$. We show that, due to strong infrared fluctuations, the latter is qualitatively similar to the corresponding one in Euclidean space $\mathbb{R}^D$ with $D=0$. It follows that spontaneously broken symmetries are radiatively restored in any space-time dimension and for any value of $N$.
\end{abstract}

\begin{keyword} 
%% keywords here, in the form: keyword \sep keyword
quantum field theory in de Sitter space \sep infrared effects\sep nonperturbative renormalization group
%% MSC codes here, in the form: \MSC code \sep code
%% or \MSC[2008] code \sep code (2000 is the default)

\end{keyword}

\end{frontmatter}
%%
%% Start line numbering here if you want
%%
%\linenumbers

%% main text
\section{Introduction}
\label{sec:intro}

The study of quantum field dynamics in de Sitter (dS) space is a subject of topical interest. It is of relevance for inflationary cosmology, where timely issues concern the understanding of radiative corrections to inflationary observables \cite{Weinberg:2005vy,Boyanovsky:2005px,Sloth:2006az,Cogollo:2008bi,Serreau:2013koa}. Moreover, the maximally symmetric dS space offers a paradigm example for the study of quantum field theory on curved space-times, which reveals a realm of nontrivial effects as compared to Minkowski space \cite{Starobinsky:1994bd,Tsamis:1996qk,Onemli:2002hr,Bros:2006gs}.
The case of light fields, with mass small in units of the curvature, is of particular interest both for phenomenological applications to inflationary cosmology and for it has no flat space analog. Such fields exhibit strong, semi-classical fluctuations on superhorizon scales, responsible for nonperturbative infrared/secular effects \cite{Starobinsky:1994bd}. Similar issues arise in various instances in flat space, e.g., for bosonic fields at high temperature or near a critical point, or for nonequilibrium systems. Powerful methods have been developed to deal with such situations, including, e.g., renormalization group, two-particle-irreducible, or large-$N$ techniques. In recent years, some efforts have been put in adapting these methods to study the  infrared (IR) dynamics of quantum fields in dS space \cite{Burgess:2009bs,Rajaraman:2010xd,Garbrecht:2011gu,Serreau:2011fu,Boyanovsky:2012qs,Youssef:2013by,Serreau:2013psa,Gautier:2013aoa}.

An interesting issue concerns the possibility of spontaneous symmetry breaking (SSB). It is known that the phase structure of a given theory can be dramatically affected by the space geometry \cite{Tarjus}. It had been argued in \cite{Ratra:1984yq}, in the case of an $O(2)$ scalar theory, that a spontaneously broken symmetry gets radiatively restored as a consequence of the large-distance logarithmic behavior of the two-point correlator of a massless field in dS space, in analogy to what happens in flat space in two dimensions \cite{Mermin:1966fe}. Similarly, a recent explicit calculation of the effective potential of $O(N)$ theories in the large-$N$ limit \cite{Serreau:2011fu} shows that strong fluctuations of superhorizon modes prevent the possibility of SSB in any space-time dimension. An simple intuitive picture may be that long range order cannot develop because of the existence of a causal horizon. If correct, this would apply to arbitrary value of $N$, including the discrete symmetry case  $N=1$.

Subsequent studies have been undertaken using either the Hartree approximation \cite{Prokopec:2011ms} or a field theoretical generalization of the Wigner-Weisskopf method \cite{Boyanovsky:2012nd}, which conclude that SSB is possible in dS space for finite $N$ and that there is a first order transition to a symmetric phase as a function of the dS radius. These studies also find that the would-be-Goldstone excitations acquire a nonzero mass. However, the Hartree approximation is known to erroneously predicts a first order transition and massive Goldstone modes in flat space \cite{Pilaftsis:2013xna}, due to the neglect of important nonlocal self-energy contributions in the broken phase \cite{Reinosa:2011ut}. This makes the analogous results in dS space dubious. Recently, employing a semiclassical stochastic approach, the authors of \cite{Lazzari:2013boa} conclude to the impossibility of SSB for arbitrary value of $N$, conforting the results of \cite{Ratra:1984yq,Serreau:2011fu} and the intuitive argument above.

In this Letter, we address the question of SSB employing nonperturbative renormalization group (NPRG) techniques \cite{Delamotte:2007pf}. In flat space, the latter have proven powerful tools in a wide variety of physical problems from condensed matter physics, to non-abelian gauge theories or to quantum gravity \cite{Gies:2006wv}. The simple local potential approximation (LPA) is able to capture the phase structure of $O(N)$ theories in arbitrary dimension, including nontrivial phenomena in two dimensions such as the Mermin-Wagner theorem for $N>2$, or, with the inclusion of an anomalous dimension, the Kosterlitz-Thouless transition for $N=2$ \cite{Delamotte:2007pf}. The LPA thus provides a unique tool to study the phase structure of $O(N)$ theories in dS space.

NPRG techniques have recently been formulated in cosmological spaces in four dimensions in \cite{Kaya:2013bga}, where the author explicitly derives flow equations in the LPA for the $N=1$ theory in the symmetric phase in dS space. Here, we generalize this derivation to arbitrary dimensions, making use of the physical momentum representation of dS correlators \cite{Busch:2012ne,Parentani:2012tx}, hereafter called the $p$-representation. The latter exploits as much as possible dS symmetries in a momentum representation and allows for a transparent formulation of the renormalization group flow directly in physical momentum space. 

We derive the explicit flow equation for the effective potential in the LPA, including a nonzero anomalous dimension, for superhorizon scales. Introducing appropriately rescaled quantities, we show that, due to strong superhorizon fluctuations, the flow equation has the same form, up to a numerical factor, as the corresponding one in Euclidean space $\mathbb{R}^D$ with $D=0$. As a consequence of this dimensional reduction, the phase structure of the $O(N)$ theory in dS space is similar to that of a zero-dimensional Euclidean theory. In particular, a broken symmetry phase is not possible whatever the dimension or the value of $N$. We illustrate this by computing the flow of the minimum of the potential and show that, under mild assumptions, the latter runs to zero in the IR. 

\section{General setup}
\label{sec:setup}

Consider the $O(N)$-symmetric scalar field theory with classical action (a sum over $a=1,\ldots,N$ is implied)
\beq
\label{eq:action}
 {\cal S}[\varphi]=\int_x\left\{{1\over2}\varphi_a\left(\square-m_{\rm dS}^2\right)\varphi_a-\frac{\lambda}{4!N}\left(\varphi_a\varphi_a\right)^2\right\},
\eeq
with the invariant measure $\int_x\equiv\int d^{D}x\,\sqrt{-g}$, on the expanding Poincar\'e patch of a $D=(d+1)$-dimensional dS space. In terms of the comoving spatial coordinates $\bX$ and the conformal time $-\infty<\eta<0$, the line-element reads (we choose the Hubble scale $H=1$)
\beq
\label{eq:metric}
 ds^2=\eta^{-2}\left(-d\eta^2+d\bX\cdot d\bX\right).
\eeq
In \Eqn{eq:action}, the mass term $m_{\rm dS}^2=m^2+\xi{\cal R}$ includes a possible coupling to the Ricci scalar ${\cal R}=d(d+1)$ and $\square$ is the appropriate Laplace operator. 

The NPRG is conveniently formulated as a flow equation for the so-called average action at the scale $\kappa$, $\Gamma_\kappa [\phi]$, which interpolates between the classical action for $\kappa\to\infty$ and the full (quantum) effective action for $\kappa\to0^+$. One introduces a regulator $R_\kappa $ as $S\to S+\Delta S_\kappa $, where
\beq
 \Delta S_\kappa [\phi]=\frac{1}{2}\int_{x,y}\phi_a(x)R_\kappa ^{ab}(x,y)\phi_b(y)
\eeq
is chosen so as to suppress fluctuations on scales below $\kappa$ while leaving high momentum modes unaffected. The average action is defined as 
$ \Gamma_\kappa [\phi]=\Gamma_\kappa ^{\rm eff}[\phi]-\Delta S_\kappa [\phi]$,
 where $\Gamma_\kappa ^{\rm eff}$ is the usual effective action corresponding to the action $S+\Delta S_\kappa $.
The evolution of $\Gamma_\kappa $ with the scale $\kappa$ is governed by the Wetterich equation \cite{Delamotte:2007pf}
\beq
\label{eq:flowGamma}
 \dot \Gamma_\kappa [\phi]=\frac{1}{2}{\rm Tr}\left\{\dot R_\kappa G_\kappa [\phi]\right\},
\eeq
where the dot stands for  $\kappa\partial_\kappa $ and the trace concerns both $O(N)$ indices and space-time variables. Here,
\beq
\label{eq:propagfull}
 G_\kappa [\phi]=i\left(\Gamma_\kappa ^{(2)}[\phi]+R_\kappa \right)^{-1}
\eeq
denotes the propagator in presence of the regulator $R_\kappa $, with $\Gamma^{(2)}_{\kappa,ab}[\phi](x,y)=[g(x)g(y)]^{-{1\over2}}\delta^2\Gamma_\kappa [\phi]/\delta\phi_a(x)\delta\phi_b(y)$.
A technical remark is in order here. Quantum field theory on cosmological spaces is con\-ve\-niently formulated as a nonequilibrium-like, initial value problem. In that case, the use of functional techniques requires one to introduce a closed  contour $\C$ in time \cite{Schwinger:1960qe}. Equations \eqn{eq:action}-\eqn{eq:propagfull} hold with the replacement $\int_x\to\int_\C dx^0\int d^dx \sqrt{-g(x)}$, where the time variable runs over the contour $\C$ \cite{Gasenzer:2008zz}.

We now formulate the flow equation \eqn{eq:flowGamma} in dS space using the $p$-representation. It is preferable to choose a regulator which respects the symmetries of the problem, here the dS and the $O(N)$ groups. However, it is also desirable to employ a momentum space description, for which the role of the regulator is transparent. The $p$-representation provides a compromise between those two demands, which are conflicting in dS space. By definition, the regulator $R_\kappa $ has the same $p$-representation as an inverse propagator. Using spatial homogeneity and isotropy in the coordinate system \eqn{eq:metric}, one writes
 \beq
 \label{eq:comf}
  R_\kappa ^{ab}(x,y)=\int\frac{d^dK}{(2\pi)^d}e^{i{\bf K}\cdot({\bf X}-{\bf X}')}\tilde R_\kappa ^{ab}(\eta,\eta',K),
 \eeq
 with ${\bf K}$ the (conserved) comoving spatial momentum. dS symmetries imply the following $p$-representation \cite{Busch:2012ne,Parentani:2012tx}
 \beq
 \tilde R_\kappa ^{ab}(\eta,\eta',K)=\left(\eta\eta'\right)^{{d+3\over2}}K^3\hat R_\kappa ^{ab}(p,p')
\eeq
where $p=-K\eta$ and $p'=-K\eta'$ are the physical momenta associated with the comoving momentum $K$ at times $\eta$ and $\eta'$ respectively. The closed time contour $\C$ can be traded for a closed momentum contour $\hat\C$ \cite{Parentani:2012tx}, on which the function $\hat R_\kappa $ is defined. Here, we employ a local, mass-like, $O(N)$-diagonal regulator
\beq
\label{eq:localreg}
 \hat R_\kappa ^{ab}(p,p')=\frac{\delta_{\hat\C}(p-p')}{p^2} \delta^{ab} R_\kappa (p),
\eeq
where the function $R_\kappa (p)$ plays the role of a heavy mass for modes $p<\kappa$. The factor $p^{-2}$ is such that a $p$-independent function $R_\kappa (p)$ would indeed correspond to a mass term \cite{Parentani:2012tx}. We emphasize that a $p$-dependent function $R_\kappa (p)$ actually breaks the full dS invariance. This is the price to pay for using a simple local regulator in physical momentum as in \eqn{eq:localreg}.  The affine subgroup of the dS group, which underlies the $p$-representation \cite{Busch:2012ne}, is left unbroken. It is important to note that this also assumes a quantum state compatible with the $p$-representation. For instance, this includes the class of $\alpha$-vacua \cite{Mottola}, but not the dS-breaking states considered, e.g., in \cite{Anderson:2013ila} in discussing the quantum stability of dS space. In the following we consider the Bunch-Davies ($\alpha=0$) state \cite{Bunch:1978yq}.

\section{Flow of the effective potential}
\label{sec:lpa}

Having in mind a derivative expansion, we write
\beq
\label{eq:ansatz}
 \Gamma_\kappa [\phi]=\int_x\left\{-V_\kappa (\phi)+{Z_\kappa \over2}\phi_a\square\phi_a\right\},
\eeq
where $V_\kappa (\phi)$ is the effective potential at scale $\kappa$ and where we included a field-strength renormalization factor $Z_\kappa $ from which one defines a running anomalous dimension $\eta_\kappa =-\dot Z_\kappa /Z_\kappa $. The LPA corresponds to $\eta_\kappa =0$ and \eqn{eq:ansatz} is sometimes called the LPA'. We stress that the fully dS invariant ansatz \eqn{eq:ansatz} is not the most general one with second order derivatives compatible with the $p$-representation. We use it here assuming that dS breaking terms are suppressed in the IR. The effective potential is defined as $\left.V_\kappa (\phi)=-\Omega^{-1}\Gamma_\kappa [\phi]\right|_{\phi={\rm const.}}$,
where $\Omega=\int_x$ is a volume factor. Writing \Eqn{eq:flowGamma} at constant field, the propagator \eqn{eq:propagfull} can be written in the $p$-representation. Introducing the comoving Fourier transform as in \eqn{eq:comf}, one has
\beq
 \tilde G_\kappa ^{ab}(\eta,\eta',K)=\frac{\left(\eta\eta'\right)^{{d-1\over2}}}{K}\hat G_\kappa ^{ab}(p,p').
\eeq
The function $\hat G_\kappa $ is defined on the closed momentum contour $\hat \C$ and can be decomposed in terms of a statistical ($\hat F_\kappa $) and a spectral ($\hat\rho_\kappa $) function as \cite{Parentani:2012tx}
\beq
\label{eq:contour}
 \hat G_\kappa ^{ab}(p,p')=\hat F_\kappa ^{ab}(p,p')-{i\over2}{\rm sign}_{\hat\C}(p-p')\hat\rho_\kappa ^{ab}(p,p'),
\eeq
where the sign function is understood on the contour. Note the symmetry properties
$\hat F^{ab}_\kappa (p,p')=\hat F^{ba}_\kappa (p',p)$ and $\hat\rho_\kappa ^{ab}(p,p')=-\hat\rho_\kappa ^{ba}(p',p)$. Finally, equal-time commutation relation imply $\partial_p\hat\rho_\kappa ^{ab}(p,p')|_{p=p'}=-\delta^{ab}/Z_\kappa $. Using Eqs. \eqn{eq:localreg}-\eqn{eq:contour} in \eqn{eq:flowGamma}, we obtain, after simple manipulations,
\beq
\label{eq:flowV}
 \dot V_\kappa (\phi)=\frac{1}{2}\int\!\frac{d^dp}{(2\pi)^d}\,\dot R_\kappa (p)\frac{\hat F_\kappa ^{aa}(p,p)}{p}.
\eeq
We note, in particular, that the volume $\Omega$ factors out in the $p$-representation.

Let us first illustrate the flow equation in the deep IR in the case $N=1$. With the ansatz \eqn{eq:ansatz}, the correlator \eqn{eq:propagfull} satisfies the  inhomogeneous equation
\beq
\label{eq:diffeq}
 \left[\partial_p^2+1-\frac{\nu_\kappa ^2-{1\over4}-Z_\kappa ^{-1}R_\kappa (p)}{p^2}\right]\!\hat G_\kappa (p,p')=\frac{\delta_{\hat\C}(p-p')}{iZ_\kappa },
\eeq
where we defined $ \nu_\kappa =\sqrt{{d^2/4}-{V_\kappa ''(\phi)/Z_\kappa }}$. 
Equivalently the functions $\hat F_\kappa $ and $\hat\rho_\kappa $ satisfy a similar equation with right hand side set to zero. The latter are solved as 
$\hat F_\kappa (p,p')=Z_\kappa ^{-1}{\rm Re}\left\{u_\kappa (p)u_\kappa ^\star(p')\right\}$ and $\hat \rho_\kappa (p,p')=-2Z_\kappa ^{-1}{\rm Im}\left\{u_\kappa (p)u_\kappa ^\star(p')\right\}$
where the function $u_\kappa $ satisfies the same homogeneous equation. Following \cite{Kaya:2013bga}, we employ the simple Litim regulator \cite{Litim:2001up}
\beq
 R_\kappa (p)=Z_\kappa (\kappa^2-p^2)\theta(\kappa^2-p^2),
\eeq
for which one has
\bea
\label{eq:vsup}
 &&\hspace{-1.1cm}u_\kappa ''(p)+\left(1-\frac{\nu_\kappa ^2-{1\over4}}{p^2}\right)u_\kappa (p)=0\quad{\rm for}\quad p\ge \kappa,\qquad\\
\label{eq:vinf}
 &&\hspace{-1.1cm}u_\kappa ''(p)-\frac{\bar\nu_\kappa ^2-{1\over4}}{p^2}u_\kappa (p)=0\quad {\rm for}\quad p\le \kappa
\eea
where $\bar\nu_\kappa =\sqrt{\nu_\kappa ^2-\kappa^2}$. 
Equation \eqn{eq:flowV} becomes
\beq
\label{eq:Vflow2}
 \dot V_\kappa (\phi)=\frac{\Omega_d}{2(2\pi)^d}\int_0^\kappa\!\!dp\,p^{d-1}\!\left[(2-\eta_\kappa )\kappa^2+\eta_\kappa p^2\right]\left|u_\kappa (p)\right|^2.
\eeq

Demanding the Bunch-Davies vacuum at large momentum \cite{Kaya:2013bga}, equations \eqn{eq:vsup} and \eqn{eq:vinf} are solved as
\bea
\label{eq:vsupsol}
 &&\hspace{-1.1cm}u_\kappa (p)=\frac{\sqrt{\pi p}}{2}e^{i\varphi_\kappa }H_{\nu_\kappa }(p) \quad{\rm for}\quad  p\ge \kappa,\\
\label{eq:vinfsol}
 &&\hspace{-1.1cm}u_\kappa (p)=\frac{\sqrt{\pi p}}{2}e^{i\varphi_\kappa }\left[c_\kappa ^-\frac{\kappa^{\bar\nu_\kappa }}{p^{\bar\nu_\kappa }}+c_\kappa ^+\frac{p^{\bar\nu_\kappa }}{\kappa^{\bar\nu_\kappa }}\right]\,\,\,\,{\rm for}\,\,\,\,  p\le \kappa,
\eea 
with $\varphi_\kappa ={\pi\over2}(\nu_\kappa +{1\over2})$ and where $H_\nu(z)$ is the Hankel function of the first kind. The continuity of $u_\kappa (p)$ and $u_\kappa '(p)$ at at $p=\kappa$ imposes
\beq
 c_\kappa ^\pm=\frac{1}{2}\left[H_{\nu_\kappa }(\kappa)\pm\frac{\kappa}{\bar\nu_\kappa }H_{\nu_\kappa }'(\kappa)\right].
\eeq
The momentum integral in \eqn{eq:Vflow2} can be performed exactly \cite{Kaya:2013bga,workinprogress}. However, we can readily obtain the flow equation in the IR regime, $\kappa\ll1$. In this Letter, we are interested in the case of light fields, for which nontrivial infrared effects come into play. More precisely, we assume small curvature $V_\kappa''(\phi)/Z_\kappa<d/2$, such that  $\nu_\kappa \in\mathbb{R}$. As explained above, this is the case of phenomenological interest for inflationary physics. Moreover, it is the relevant case for the study of the phase structure of $O(N)$ theories in dS space. Indeed, consider the case of a SSB potential at a UV scale $\kappa\gg1$. Integrating out subhorizon momenta generates the standard Minkowski flow which brings the effective potential to a very flat shape for scales $\kappa\sim1$ \cite{Delamotte:2007pf}. Here, we wish to study the further evolution of such a flat potential as one integrates out superhorizon scales. Finally, we mention that the IR finiteness of the flow equation \eqn{eq:Vflow2} requires $\bar\nu_\kappa<d/2$, that is $\kappa^2+V_\kappa''(\phi)/Z_\kappa>0$.

Using $H_\nu(z)=2^\nu\Gamma(\nu)/ (i\pi z^{\nu})[1+{\cal O}(z^2)]$ and the definition of $\bar\nu_\kappa $, we find, up to relative corrections of ${\cal O}(\kappa^2)$, $c_\kappa ^-=2^{\nu_\kappa }\Gamma(\nu_\kappa )/(i\pi \kappa^{\nu_\kappa })$ and  $ c_\kappa ^+=-(\kappa/2\nu_\kappa )^2c_\kappa ^-$. One easily checks that the contribution to \eqn{eq:Vflow2} from $c_\kappa ^+$ in \eqn{eq:vinfsol} is thus IR suppressed. After some simple calculations we obtain, up to relative corrections of ${\cal O}(\kappa^2)$,
\beq
\label{eq:Vflow3}
  \dot V_\kappa (\phi)=\frac{\Omega_dF_{\nu_\kappa }}{2(2\pi)^d}\kappa^{d+2-2\nu_\kappa }\left\{\frac{2-\eta_\kappa }{d-2\bar\nu_\kappa }+\frac{\eta_\kappa }{d+2-2\bar\nu_\kappa }\right\},
\eeq 
 with $F_{\nu}=[2^\nu\Gamma(\nu)]^2/(4\pi)$. Here, the trivial dimensional factor $\kappa^{d+2}$ gets modified by a large dynamical contribution  $\sim\kappa^{-2\nu_\kappa }$ from enhanced IR fluctuations.
 
\begin{figure}[t!]  
\epsfig{file=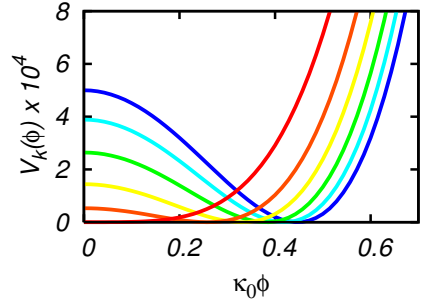,width=6.cm}
 \caption{\label{fig:flot} 
The flow of the potential in the LPA ($\eta_\kappa =0$) with the  ansatz $V_\kappa (\phi)=\lambda_\kappa (\rho-{\bar\rho}_\kappa  )^2/2$, where $\rho=\kappa^2\phi^2/2$, for $N=1$ and $d=3$ ($A_3=3/16\pi^2$). The initial parameters at the arbitrary scale $\kappa_0\lesssim1$ are $\lambda_{\kappa_0}=\bar\rho_{\kappa_0}=0.1$. The curves from right to left correspond to $\ln \kappa_0/\kappa=0,0.1,0.2,0.3,0.4,0.51$. The potential flattens as its minimum decreases and reaches zero at a finite scale (leftmost curve).}
\end{figure}

To obtain a form of the flow equation with no explicit dependence on $\kappa$, we introduce the rescaled variable
\beq
\label{eq:rescale}
 \rho=\frac{Z_\kappa }{2}\kappa^2\phi^2 \quad{\rm and}\quad U_\kappa (\rho)=V_\kappa (\phi), 
\eeq 
such that $V_\kappa ''(\phi)= \kappa^2[U_\kappa '(\rho)+2\rho U_\kappa ''(\rho)]$. We thus have $d-2\nu_\kappa ={\cal O}(\kappa^2)$ and $d-2\bar\nu_\kappa \approx (2/d)\kappa^2[1+U_\kappa '(\rho)+2\rho U_\kappa ''(\rho)]$ and  the first term in bracket in \eqn{eq:Vflow3} gets further IR enhanced by a factor $\kappa^{-2}$. We obtain, in the limit $\kappa\ll1$,
\beq
\label{eq:Vflow4}
 \dot U_\kappa (\rho)=(2-\eta_\kappa )\left\{-\rho U_\kappa '(\rho)+\frac{A_d}{1+U_\kappa '(\rho)+2\rho U_\kappa ''(\rho)}\right\},
\eeq
with $A_d=d\Omega_dF_{d/2}/4(2\pi)^d=d\Gamma(d/2)/8\pi^{d/2+1}$ and where the prime denotes a derivative with respect to $\rho$. As announced, the flow equation \eqn{eq:Vflow4} is similar to the corresponding one in Euclidean space $\mathbb{R}^D$ with $D=0$  \cite{Delamotte:2007pf}, up to the factor $A_d$. One thus expects the phase structure in dS space to be qualitatively the same as that of the flat Euclidean theory in low dimension. 

The minimum ${\bar\rho}_\kappa  $ of $U_\kappa (\rho)$ is defined as $U_\kappa '({\bar\rho}_\kappa  )=0$. Using $\dot U_\kappa '({\bar\rho}_\kappa  )+\dot{\bar\rho}_\kappa   U''_\kappa ({\bar\rho}_\kappa  )=0$, one gets the flow equation
\beq
\label{eq:flowmin1}
 \dot{\bar\rho}_\kappa  =(2-\eta_\kappa )\left\{{\bar\rho}_\kappa  +A_d\frac{3+2{\bar\rho}_\kappa  g_\kappa }{(1+2{\bar\rho}_\kappa  \lambda_\kappa )^2}\right\},
\eeq
where $ \lambda_\kappa =U''_\kappa ({\bar\rho}_\kappa  )\ge0$ and $g_\kappa \lambda_\kappa =U'''_\kappa ({\bar\rho}_\kappa  )$. Observe that $\dot{\bar\rho}_\kappa  >0$ under reasonable assumptions on $\eta_\kappa $ and $g_\kappa $. In that case, ${\bar\rho}_\kappa  $ decreases in the IR and reaches zero at a finite renormalization scale. The physical minimum $\bar\phi_\kappa $, defined as ${\bar\rho}_\kappa  =Z_\kappa \kappa^2\bar\phi_\kappa ^2/2$, also runs to zero under conservative assumptions concerning the flow of $\eta_\kappa $, e.g., $\eta_{\kappa\to0}\to{\rm const}$.
This is illustrated in Fig. \ref{fig:flot}, where \Eqn{eq:Vflow4} is solved in the LPA with a quartic ansatz $U_\kappa (\rho)=\lambda_\kappa (\rho-{\bar\rho}_\kappa  )^2/2$, for which $\eta_\kappa =g_\kappa =0$.

The previous analysis is easily extended to $N>1$. One has to distinguish longitudinal and transverse modes according to the projectors $P^L_{ab}=\phi_a\phi_b/\phi^2$ and $P^T_{ab}=\delta_{ab}-P^L_{ab}$. In particular, the curvature of the potential now becomes, in terms of the definitions \eqn{eq:rescale}, $V_\kappa ''(\phi)\to \kappa^2[U_\kappa '(\rho)+2\rho U_\kappa ''(\rho)]P^L_{ab}+\kappa^2U_\kappa '(\rho)P^T_{ab}$. After similar manipulations as before, we get, in the limit $\kappa\to0$,
\beq
\begin{split}
 \dot U_\kappa (\rho)&=(2-\eta_\kappa )\bigg\{\!-\rho U_\kappa '(\rho)\\
 \label{eq:flowNN}
 &+A_d\left[\frac{N-1}{1+U_\kappa '(\rho)}+\frac{1}{1+U_\kappa '(\rho)+2\rho U_\kappa ''(\rho)}\right]\bigg\}.
\end{split}
\eeq
As before, this equation is similar to the corresponding one in flat Euclidean space with $D=0$ \cite{Delamotte:2007pf} and the flow drives the system to the symmetric phase under mild assumptions on $\eta_\kappa $ and $g_\kappa $. For instance, \Eqn{eq:flowmin1} becomes
\beq
\label{eq:flowminN}
 \dot{\bar\rho}_\kappa  =(2-\eta_\kappa )\left\{{\bar\rho}_\kappa  +A_d\left[N-1+\frac{3+2{\bar\rho}_\kappa  g_\kappa }{(1+2{\bar\rho}_\kappa  \lambda_\kappa )^2}\right]\right\}\!.\,\,
\eeq

We conclude that the NPRG flow equations in the LPA' generically predict that spontaneously broken $O(N)$ symmetries are radiatively restored in dS space in any dimension and for all values of $N$. This results from an effective dimensional reduction due to strong superhorizon fluctuations.\footnote{It is interesting to note that a similar dimensional reduction phenomenon, with $d+1\to1+1$, has been observed for fermionic degrees of freedom in spaces of constant negative curvature; see, e.g.,  \cite{Gorbar:1999wa,Gies:2013dca}.} Our findings support the previous results of Refs. \cite{Ratra:1984yq,Serreau:2011fu,Lazzari:2013boa}. This assumes, in particular, a smooth and small enough 
 anomalous dimension $\eta_\kappa $. There are examples in flat space where this is not the case, such as for the Kosterlitz-Thouless transition \cite{Delamotte:2007pf}. It is of great interest to investigate this possibility in the present case \cite{workinprogress}. Also of interest is the (numerical) study of the transition from the ultraviolet, where one should recovers the Minkowski flow, to the IR where strong superhorizon fluctuations come into play. Finally, it is important to study the role of possible dS symmetry breaking terms allowed by the $p$-representation along the flow.

\section*{Acknowledgements}

I thank M. Tissier and N. Wschebor for interesting discussions and M. Tissier for useful remarks concerning the manuscript.

\end{document}